\documentclass{emulateapj}
\usepackage{lscape}

\slugcomment{\today}

\shorttitle{BAL Quasar 1045+352}
\shortauthors{M. Kunert-Bajraszewska et al.}

\begin{document}

\title{Origin of the complex radio structure in BAL\,QSO 1045+352}

\author{Magdalena Kunert-Bajraszewska$^{1}$, Agnieszka Janiuk$^{2}$, Marcin P. Gawro\'nski$^{1}$, 
Aneta Siemiginowska$^{3}$}

\affil{$^{1}$ Toru\'n Center for Astronomy, N. Copernicus University,
Gagarina 11, 87-100 Toru\'n,, Poland}
\affil{$^{2}$ N. Copernicus Astronomical Center, Bartycka 18, 00-716
Warsaw, Poland}
\affil{$^{3}$ Harvard Smithsonian Center for Astrophysics, 60 Garden St, 
Cambridge, MA 02138}

\begin{abstract}

We present new more sensitive high-resolution radio observations of
a compact broad absorption line (BAL) quasar, 1045+352, made with the EVN+MERLIN at 5 GHz.
They allowed us to trace the connection between the arcsecond
structure and the radio core of the quasar.  The radio morphology of
1045+352 is dominated by a knotty jet showing several bends.
We discuss possible scenarios that could explain
such a complex morphology: galaxy merger, accretion disk instability,
precession of the jet and jet-cloud interactions.
It is possible that we are witnessing an ongoing 
jet precession in this source due to internal instabilities within the jet flow,
however, a dense environment detected in the submillimeter band
and an outflowing material suggested by the X-ray absorption 
could strongly interact with the jet.  It is difficult to
establish the orientation between the jet axis and the observer in
1045+352 because of the complex structure. Nevertheless taking into
account the most recent inner radio structure we conclude that the radio
jet is oriented close to the line of sight which can mean that the
opening angle of the accretion disk wind can be large in this source.
We also suggest that there is no direct correlation between
the jet-observer orientation and the possibility of observing BALs.
\end{abstract}


\section{Introduction}
Broad absorption lines (BALs) - high ionization
resonant lines (C\,IV 1549$\AA$) and low ionization lines (Mg\,II
2800$\AA$), are seen in a small number (15\,\%) of both the
radio-quiet and radio-loud quasar populations \citep{dai, knigge, gibson}, 
according to the traditional
BAL\,QSO definition of \citet{weymann91}. They are probably caused
by the outflow of gas with high velocities and are part of the accretion
process. The presence of BALs is a geometrical
effect \citep{elvis00} and/or is
connected with the quasar evolution \citep{becker00,gregg00, gregg06}.

Theoretical models \citep{elvis00,murray95} suggest that BALs
are seen at high inclination angles, which means that the outflows
from accretion disks are present near the equatorial
plane. However, some recent numerical work indicates
that it is also plausible to launch bipolar outflows from
the inner regions of a thin disk \citep[e.g.][]{ghosh07,prog04}.
There is growing observational evidence indicating
the existence of polar BAL outflows \citep[e.g.][]{zhou06,ghosh08}.
This means that there is no simple orientation model which can explain
all the features observed in BAL quasars.

It has been suggested by \cite{becker00} that, because of their small
sizes, most of the radio-loud BAL quasars belong to the class of
compact radio sources, namely compact steep spectrum (CSS) objects and
gigahertz peaked spectrum (GPS) objects. GPS and CSS sources are
considered to be young radio sources with linear sizes less than
20\,kpc, entirely contained within the extent of the host galaxy.  The
high resolution observation in the VLBI technique is the best way to
learn about their morphologies and orientation. However, very small
fraction of compact BAL quasars have been observed with VLBI so far
\citep{jiang03,kun07,liu08,monte09, monte08}.  About half of them still have
unresolved radio structures even in the high resolution
observations, the others have core-jet structures indicating some
re-orientation or very complex morphology, suggesting a strong
interaction with the surrounding medium. All of them are potentially
smaller than their host galaxies. The analysis of the spectral shape,
variability and polarization properties of some of them shows that
they are indeed similar to CSS and GPS objects, and are not oriented
along a particular line of sight \citep{monte09}.

1045+352 is a compact radio-loud CSS source.  The spectral
observations \citep{willott02} have shown 1045+352 to be a quasar with
a redshift of $z=1.604$. It has also been classified as a HiBALQSO
based upon the observed very broad C\,IV absorption, and it is a very
luminous submillimeter object with detections at both 850\,$\mu$m and
450\,$\mu$m \citep{willott02}.  The radio luminosity of 1045+352 at
1.4\,GHz is high (log$L_{1.4\mathrm{GHz}}$=28.25~W~${\rm Hz^{-1}}$), making
this source one of the most radio-luminous BAL quasars, with a value
similar to that of the first known radio-loud BAL\,QSO with an FR\,II
structure \citep{gregg00}.  The {\it Chandra} X-ray observation of 1045+352
\citep{kunert09} shows that the X-ray emission of 1045+352 is very weak
in comparison to the other radio-loud BAL quasars which, together with
the high value of the optical$-$X-ray index $\alpha_{ox}=1.88$,
suggests the presence of an X-ray absorber close/in the BLR region.  The
result of the spectral energy distribution (SED) modeling of 1045+352 indicate
that the X-ray emission we
observe from 1045+352 may mostly be due to X-ray emission from the
relativistic jet, while the X-ray emission from the corona is absorbed
in a large part.  The first multifrequency radio observations of the
quasar 1045+352 were made by \citet{kun07}.  The complex compact
structure has been resolved into many subcomponents and indicates
that the jet is moving in a non-uniform way in the central regions of
the host galaxy.  Here we present new, more sensitive, high resolution 
EVN+MERLIN 5\,GHz observations of this source and analyze them.

\section{Radio Observations}
\label{obs}

1045+352 belongs to the primary sample of 60 candidates for CSS
sources selected from the VLA FIRST catalogue \citep{wbhg97}. Initial
observations of all the candidates were made with MERLIN at 5\,GHz
\citep{mar03} and 1.7, 5 and 8.4\,GHz VLBA follow-up of 1045+352 was
carried out on 13 November 2004 in a snapshot mode with
phase-referencing \citep{kun07}.  The observations showed that this
source has a complex radio morphology with a probable radio jet axis
reorientation.  To look for the traces of the above-mentioned scenario
we planned more sensitive high-resolution EVN+MERLIN 5 GHz observation
that was carried out on 2 June 2007 in a full-track mode.  
The target source itself was observed for ~7\,hrs.
The whole data reduction process was carried out using standard AIPS
procedures. IMAGR was used to produce the final
images separately for EVN and MERLIN, and finally the combined
EVN+MERLIN image was created (Fig.~\ref{1045+352_maps}).  The flux
densities of the main components of the target source were then
measured, by fitting Gaussian models, using task JMFIT
(Table~\ref{fluxes}). For more extended features the flux densities
were evaluated by means of IMSTAT (*, Table~\ref{fluxes}). The basic parameters of
1045+452 have been gathered in Table~\ref{table1}.

Throughout the paper, we assume the cosmology with
${\rm
H_0}$=71${\rm\,km\,s^{-1}\,Mpc^{-1}}$, $\Omega_{M}$=0.27,
$\Omega_{\Lambda}$=0.73.

\begin{table}[t]
\begin{center}
\caption{Basic parameters of 1045+352}
\begin{tabular}{@{}l r@{}}
\hline
\hline
Parameter & Value\\
\hline
Source name (B1950)    & 1045+352\\
RA (J2000) extracted from FIRST & $10^{\rm h}~48^{\rm m}~34\fs247$\\
Dec (J2000) extracted from FIRST & $+34^{\rm o}~57\farcm0~24\farcs99$\\
Redshift {\it z}& 1.604\\
Total flux density $S_{1.4\,{\rm GHz}}$~(mJy)&1051 \\
log$L_{1.4\mathrm{GHz}}$~(W~${\rm Hz^{-1}}$)& 28.25\\
Total flux density $S_{4.85\,{\rm GHz}}$~(mJy)& 439 \\
Spectral index $\alpha_{1.4\mathrm{GHz}}^{4.85\mathrm{GHz}}$&$0.70$\\
Largest Angular Size measured&  \\
in the 5\,GHz MERLIN image&$0\farcs5$\\
Largest Linear Size ($h^{-1}~{\rm kpc}$) & 4.3\\
Radio-loudness parameter $R^{*}$ & 4.9(4.1)\\
Intrinsic extinction $A_{V}$ & 1.5\\
K-corrected absolute magnitude $M_{V}$ & -22.83(-24.33)\\ 
\hline
\end{tabular}
\label{table1} 
\end{center}
Notes.Quantities in parentheses are corrected for intrinsic extinction.
Spectral index defined as ($S\propto\nu^{-\alpha_{r}}$)
\end{table}


\section{Results}
\label{results}

\begin{table}[t]
\begin{center}
\caption{Flux densities of 1045+352 principal components from the 5\,GHz
MERLIN and EVN images}
\begin{tabular}{ccccc}
\hline
\hline
Compo- & {\small MERLIN} & {\small EVN} & {\small EVN}&$\theta_{maj}\times \theta_{min}$\\
nents  &      &   & {\small +}& \\
       &      &   & {\small MERLIN}& \\
       &   mJy  & mJy & mJy&arcs \\
\hline
A+C &$-$ &$104.1\pm 0.4$&$-$& $0.01\times 0.004$ \\
$A_{1}$ &$-$ &51(*)&$-$& $-$ \\
$A_{2}$ &$-$ &$-$&20(*)& $-$\\
$A_{3}$ &$6.0\pm 0.2$ &$-$&$-$& $0.06\times 0.01$ \\
B &$-$ &$3.8\pm 0.2$& $-$& $0.006\times 0.001$ \\
\hline
\end{tabular} 
\label{fluxes}
\end{center}
Note. The asterix shows the sum of the visible components.
\end{table}

Our previous MERLIN image of 1045+352 \citep{kun07} shows this
source to be extended in both NE/SW and NW/SE directions suggesting a
reorientation of the jet axis.

\begin{figure*}[t]
\centering
\includegraphics[width=18cm, height=18cm]{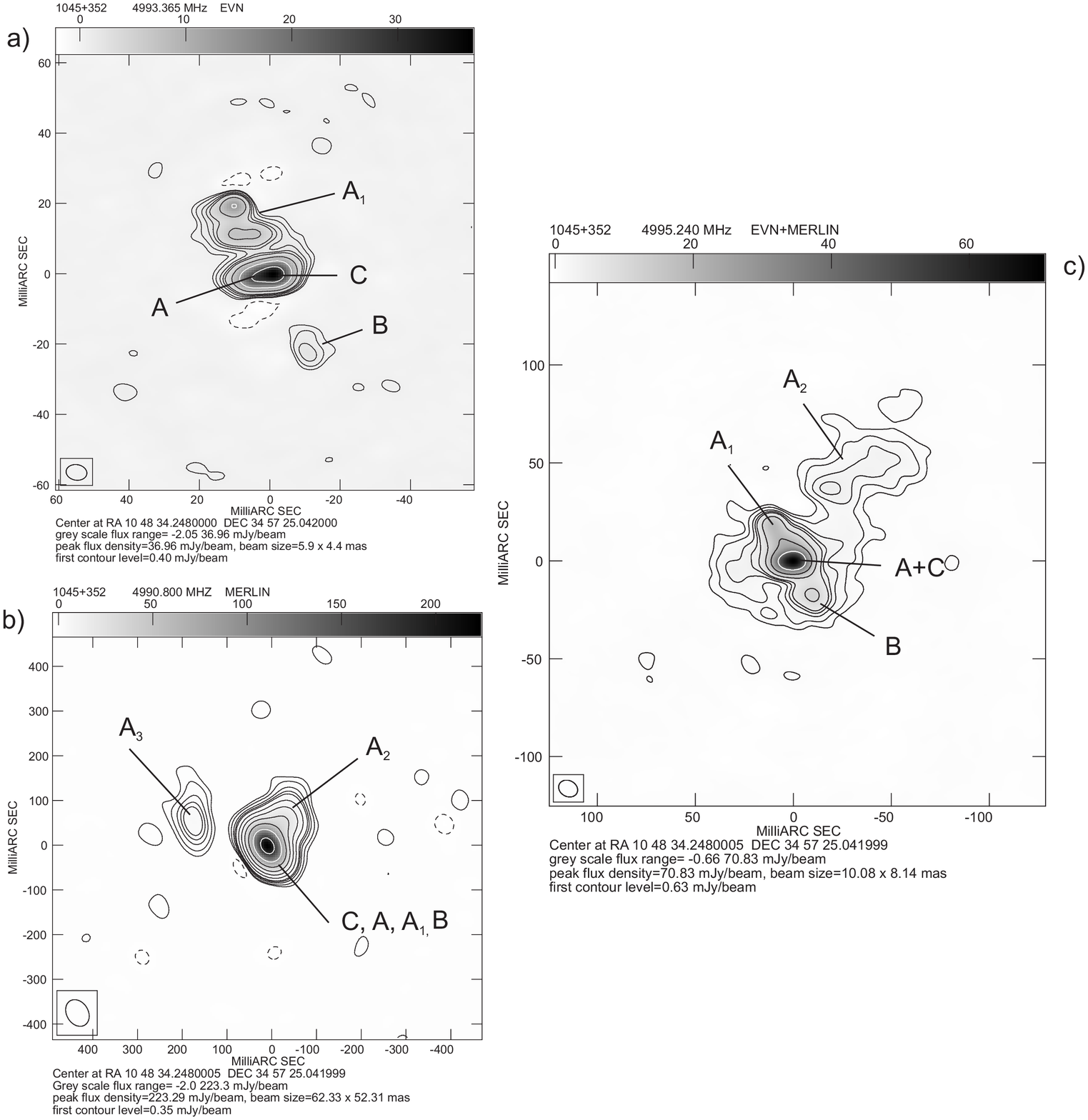}
\caption{Radio images of 1045+352: a) 5\,GHz EVN image, 
b) 5\,GHz MERLIN image, c) combined EVN+MERLIN 5\,GHz
image. Contours increase
by a factor 2 and the first contour level corresponds to $\approx 3\sigma$.}
\label{1045+352_maps}
\end{figure*}

The new EVN+MERLIN 5\,GHz radio observations of 1045+352 BAL quasar
revealed more details of its complex morphology
(Fig.~\ref{1045+352_maps}). We identify the
structure C with the radio core and the structures A - $A_{3}$ with the
subsequent jet activity.
The new 5\,GHz EVN observation shows a
jet (indicated as a A, Fig.~\ref{1045+352_maps}a) emerging from the core
(C) in the E/SE direction, and then another jet ($A_{1}$) emerging in
the NE direction. The feature indicated as B is probably a
counter-jet, and its trace was also visible in the previous lower
resolution 1.7\,GHz VLBA image \citep{kun07}. No corresponding
$B_{1}$-$B_{3}$
structures are found, possibly due to their weakness.
There is an appreciable ($\sim60^{\rm o}$) misalignment between the axis of
the inner structure (jet A) with respect to the outer structure (jet
$A_{1}$). The new combined image
shows diffuse radio emission in the center of 1045+352 quasar
suggesting very dense medium of the host galaxy of the source and
indicating strong interactions (Fig.~\ref{1045+352_maps}c).

The projected linear size of the whole structure measured in our old
snapshot 5\,GHz image \citep{kun07} as a size of the contour plot is
4.3\,kpc (Table~\ref{table1}). The projected
distance between the radio core C and the feature $A_{3}$ in the new 5\,GHz
MERLIN image (Fig.~\ref{1045+352_maps}b) measured as a separation
between the component peaks is 1.7\,kpc.  The projected size of the
innermost part (A+C) is 190\,pc. The size of the radio core amounts to
68\,pc when measured in the 8.4\,GHz VLBA image \citep{kun07} where the
innermost components were resolved.  The measured component fluxes
from the 5\,GHz EVN and MERLIN images are collected in
Table~\ref{fluxes}.

Based on the higher resolution VLBA image \citep{kun07} and using the
radio-to-optical luminosity ratio \citep{wills95} we have estimated
the angle between the jet axis and the line of sight to be
$\theta\sim30\degr$.  However, this should be treated as a rough
approximation since the radio-to-optical luminosity ratio
\citep{wills95} was defined for large scale objects and itself
suffers from very large uncertainty.  In addition, the relation we used is
based on the radio luminosity of the core, so we estimate that the
derived value of the angle corresponds to the innermost part of the
jet - component A.

Component $A_{3}$ (Fig.~\ref{1045+352_maps}b) seems to be the oldest
structure of 1045+352. A new emission - radio jet A is emerging from
the core C
and an  earlier phase of the jet activity is represented by
components $A_{1}$ and B.  Because of their symmetry with respect to the
radio core, we treated them as jet and counter-jet, respectively. The
{\sl jet/counter-jet} brightness ratio is given by:

\begin{equation}
R=\left(\frac{1+\beta cos \Theta}{1-\beta cos
\Theta}\right)^{2+\alpha_{jet}}
\end{equation}

where $\beta$ is the jet/counter-jet speed, $\theta$ is an angle
between the jet axis and the line of sight and $\alpha_{jet}$ is the
jet average spectral index.  
The value of the sideness parameter is $\rm R\sim13.5$.
Then assuming a typical value of the spectral index
$\alpha_{jet}=0.75$ and the jet speed $\beta=0.9\,c$ we obtained the
value of the angle between the jet axis and the line of sight to be
$\theta\sim61\degr$. However, $\theta$ varies in the range $\rm
51\degr-64\degr$ for $\beta$ and $\alpha_{jet}$ values in the range
$\rm 0.7\,c-0.98\,c$ and $\rm 0.7-1.1$ respectively. This may indicate
a large change between the orientation of 1045+352 radio structures on the
different scales and fits well into the picture of complicated radio
morphology.

An assumption of $\theta=30\degr-61\degr$ yields the deprojected linear
size of the source of $\sim5-9$~kpc, also indicating a young radio
source.

\section{Discussion}
\label{dis}

The complex morphology and a change of the jet direction observed in
1045+352 may result from (1) a merger, (2) a jet precession, or (3)
jet-cloud interactions.  Below, we discuss possible origins of the
multiple radio structures and their misalignment.

\subsection{Quasar reactivation}
\subsubsection{Galaxy merger}

Observations indicate, that about 50\% of young AGN contain double
nuclei in their host galaxies or exhibit morphological distortions
that are supposed to be due to the past merging events
\citep{odea98,liu04}.  
Although there are no obvious signs of merger event in the optical image of
1045+352 (from the Sloan Digital Sky Survey, SDSS), the radio distortions 
indicate this possibility.
Therefore we discuss below the possible scenario of a merger event 
in QSO 1045+352.
As shown by Liu (2004), the spin axis of the black hole formed
after the merger changes its orientation from the vertical with
respect to the outer accretion disk to the aligned with the rotation
axis of the binary on timescale $10^{5}$ yr.

The
angle by which the jet might have precessed in 1045+352 is difficult
to be determined observationally, due to orientation effects.  The
inclination angle of the A+C structure (jet axis) to the line of sight
is about $30^{\circ}$. In the case of the older components $A_{1}$ and B it
seems to be rather $\theta\sim61\degr$.  The images of the radio
structures are projected onto the plane of the sky.  Therefore even
for a very small precession angle, e.g.  $10^{\circ}$, the misaligned
structure will appear shifted by a large angle on the 2-D map, e.g.
$90^{\circ}$, due to the projection effect.

In the case of 1045+352, a timescale for the
most distant structure $A_{3}$ ($1.7$ kpc, in projection) will be equal to
$t=5.5\times 10^{4}$ yrs if we assume the jet velocity of $0.1 c$, and
$t=1.8\times 10^{4}$ yrs for $0.3 c$.  Therefore in this source, we
may be witnessing an ongoing process of the disk realignment after a
merger event, which took place after structure $A_{3}$ was ignited.

\begin{figure}[!t]
\includegraphics[width=9cm, height=9cm]{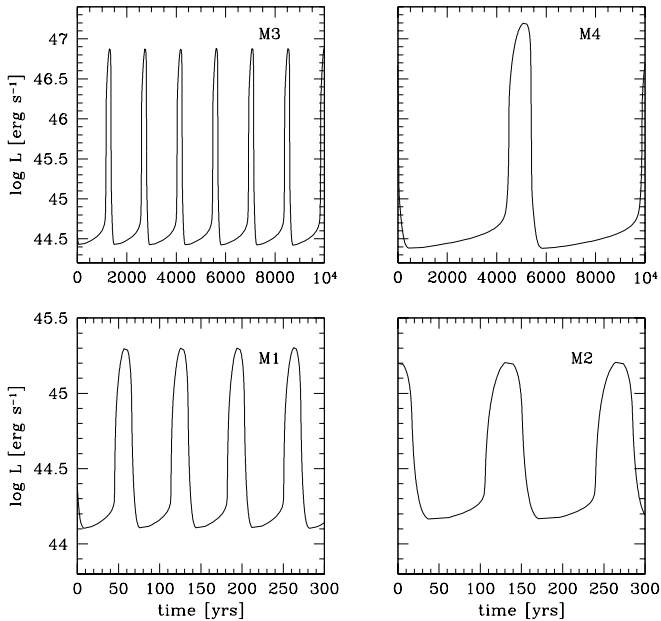}
\caption{The time evolution of the core luminosity for 2 values of mass and
the mean accretion rate and 2 values of the viscosity parameter: $\dot
m=0.36,
0.68$, $M = 2\times 10^{7}, 5.6\times 10^{7} M_{\odot}$,
and $\alpha=0.03, 0.1$. The parameters are given in the
Table~\ref{tab:evol_models}.}
\label{fig:evol_1}
\end{figure}

\begin{table}[!t]
\caption{Models of the accretion disk evolution}
\begin{tabular}{lrrrrr}
\hline
\hline
Model & $\alpha$ & $\dot m$ [$\dot M_{\rm Edd}$] & $M [M_{\odot}]$ & $T_{\rm
q}$ [yrs] & $ T_{\rm out}$ [yrs] \\
\hline
M1 &    0.1   & 0.36    & $2\times 10^{7}$  &   45     &     25 \\
M2 &    0.03  & 0.36    & $2\times 10^{7} $ &   85     &     45 \\
M3 &    0.1   & 0.68    & $5.6\times 10^{7} $ & 1.75$\times 10^{3}$     &
235         \\
M4 &    0.03  & 0.68    & $5.6\times 10^{7} $ & 4.35$\times 10^{3}$     &
880         \\
\hline
A1 &    0.01 &  1.0     & $4\times 10^{8} $ &   8.5$\times 10^{4}$      &
2.05$\times 10^{4}$ \\
A2 &    0.01 &  0.7     & $4\times 10^{8} $ &   1.5$\times 10^{4}$      &
8.1$\times 10^{3}$ \\
B1 &    0.1  &  1.0     & $4\times 10^{8} $ &   3.8$\times 10^{4}$      &
4.8$\times 10^{3}$ \\
B2 &    0.1  &  0.7     & $4\times 10^{8} $ &   2.5$\times 10^{4}$      &
1.8$\times 10^{3}$ \\
\hline
\end{tabular}

Totes. $\alpha$ - viscosity parameter, $T_{\rm q}$ - quiescent time, $T_{\rm out}$
- outburst time.
\label{tab:evol_models}
\end{table}

\subsubsection{Accretion disk instability}

In this section, we apply the accretion disk instability model to
explain the "reactivation" of the 1045+352 core.  Below, we describe in
brief the basic properties of this model, which was described in more
detail elsewhere \citep{jan02,czerny}.

Instabilities of the accretion disks have been shown to operate on
long \citep{siemi96,jan04,janiuk} and short timescales
\citep{jan02,czerny} , depending on their nature 
(the dwarf-nova type of instability,
caused by the partial ionization of hydrogen; 
or the radiation pressure instability, studied
in microquasars).
Scaling the outbursts of the Galactic microquasar GRS 1915+105,
lasting about 100$-$2000 seconds for a black hole mass of $10
M_{\odot}$, to the quasar hosting a supermassive black hole,
gives outbursts on timescales of $10^{2}-10^{5}$ years.

The unstable disk surrounding a black hole is
subject to thermal and viscous instability if the radiation
pressure dominates over the gas pressure.  
If the accretion rate outside the unstable region
(i.e. the mean accretion rate) is such that the disk is in the
unstable mode, the source enters a cycle of bright, hot states,
separated by cold, quiescent states. In the hot state, the
luminous quasar could power a radio jet,
while during the cold state the radio activity ceases.

The quantitative results, i.e. the outburst amplitudes and durations,
are sensitive to the model parameters: black hole mass and viscosity
coefficient.  In addition, the description of the heating is essential. If
the heating is proportional to the total pressure, the outburst
amplitudes are very large.  The heating proportional to the square
root of the gas times the total pressure reduces the amplitude of the
disk outbursts. 
  
We compute the time dependent accretion disk model for several
combinations of the main parameters: four values of the mean accretion
rate, $\dot m=0.36, 0.68, 0.7$ and 1.0 in units of the Eddington
accretion rate; three values of the viscosity parameter: $\alpha=0.01,
0.03$ and 0.1; and three values of the black hole mass, $M= 2\times
10^{7}, 5.6\times 10^{7}$ and $4\times 10^{8} M_{\odot}$ 
(see Table 3; models are labeled with A, B, and M).
The
accretion rate is in the units of the Eddington accretion rate, $\dot
M_{\rm Edd}$, which corresponds to the Eddington luminosity.  In
Table \ref{tab:evol_models} we give the resulting time for the
quiescent and outburst periods. In Figure \ref{fig:evol_1} we show the
exemplary light curves from the time dependent accretion disk model. In
general, the outburst separation is consistent with the one required
by \citet{rb97}, who proposed a phenomenological model to fit the
observed number of radio sources and their sizes.

In our model, for a given black hole mass, the larger the
mean accretion rate, the longer the duration of a cycle episode, both
in hot and cold states. In addition the $\alpha$ parameter affects the
results and for smaller viscosity, the cycle duration is longer. This
relation was calibrated by \citet{czerny} using a grid of models for
various black hole masses and Eddington ratios. The proper value of a
bolometric correction in this relation can give us an estimation of an
upper limit for the age of a radio source (i.e. the duration of the
hot state).  The predicted timescales and durations are in agreement
with observations of compact radio sources \citep{wu}.

The outbursts are associated with the ejections of radio jets.  The
jets are then turned-off between the outbursts and each radio
structure will represent a new outburst. In case of an apparently
young, compact source we can suspect that in fact it is an old,
reactivated object, in which the vast radio structures have already faded
away and are not visible. This mechanism would explain the
apparent statistical excess of the compact sources with respect to the
galaxies with extended radio structures (\citet{odea97}).  On the
other hand, the timescales for fading of a radio source may be much
longer than the separation timescales between the outbursts and we
should observe the fossil radio structures in addition to the compact
one \citep{baum90}.  These kind of morphologies have already been observed in large
scale objects showing evidence of two or more active periods
\citep[e.g][]{schoe, jamrozy}.

Here, we present the results of the modeling of the intermittent
activity for the source 1045+352. The best fitting parameters for the
quasar 1045+352 from \citet{kunert09} were $M_{BH}=2 \times 10^{7}
M_{\odot}$, $\dot m=0.36$ and $\alpha=0.1$ (model M1).  
Such a large accretion
rate is high enough for the radiation pressure instability to operate
($\dot M_{\rm crit} \approx 0.2 M_{\rm Edd}$).  We take these
parameters as an input to our time-evolution computations,
and for comparison we calculate model M2, with a different viscosity.
The accretion rate in solar masses per year is therefore $\dot M=\dot m
\times 3.52\,M_{8}= 0.253\,M_{8}$, where $M_{8}$ is the black hole
mass in the units of $10^{8}$ Solar masses.  Because the black hole
mass determination for this source is not very certain, and seems to
be rather a lower limit, we calculated also several models for
somewhat larger masses (models M3, M4, for which we used the mass estimation 
from the monochromatic luminosity; models A and B, where the assumed mass 
value enables the radio source to escape from the host galaxy).  
The model is quite sensitive for the adopted
central mass value and the results presented here are only examples.
Full parameter space for the mass and accretion rates for large sample
of quasars was presented in \citet{czerny}.

In the bottom left panel of Fig. \ref{fig:evol_1} we plot the
light curve for the given best fit parameters, taken from the spectral
modeling \citep{kunert09}.  The parameters are given in the Table
\ref{tab:evol_models}, model M1.  The resulting duration of an active
phase in this case would be about 25 years, while for the quiescent
phase it is about 45 years \citep[see][]{czerny} for the dependence
between the activity timescale and bolometric luminosity of the
source).

In this scenario, the subsequent radio structures would be fed by the jet
produced during the active phase of the central engine. In the quiescent
phase, the jet stops feeding the radio lobe, and further expansion of the
remnant radio structure proceeds due to the accumulated jet energy.
If the proper motion of the outer jet components in 1045+352 is in the order
of $0.1-0.2\,c$, and the restart of the activity was after $T_{\rm q}\sim$
50 yrs, then the physical separation of the radio structures would be of about
10\,pc. 
If the velocities of the jet were rather of $0.7-0.9\,c$, such as those characteristic
of the inner structures $A_{1}$ - B, then this separation would increase to
35$-$45\,pc.

During the active phase, the radio lobes are fed by the expanding jet
and the source grows. For the outburst duration calculated in model
M1, the size of the central radio core would be only about 1.5\,pc for
a jet velocity of $0.1\,c$, 4.6\,pc for velocity of $0.3\,c$ and
15\,pc for a relativistic jet speed of $0.9\,c$. The resulting size of
the radio source is therefore too small in comparison with the
structure size estimated from our images (the size of the radio core is
$\sim$70\,pc). The much larger actual size requires longer activity
timescales, which are possible in the models with larger black hole
mass and accretion rate and/or smaller viscosity. We therefore
checked
a number of additional models, with larger black hole mass and
accretion rate.  In particular, the subsequent models M2, M3 and M4
adopt these parameters within the range of the plausible values for
our previous spectral modeling.  These values were not the best fit,
however, they also gave a reasonable shape of the high energy spectrum
of this quasar and therefore cannot be excluded.  As Table
\ref{tab:evol_models} shows, model M4, with black hole mass of
$5.6\times 10^{7} M_{\odot}$, accretion rate of $\dot m = 0.68$ and
viscosity $\alpha = 0.03$ results in an outburst timescale of almost
900 years, which gives a size of the radio core of 30$-$80 pc,
depending on the jet velocity (non-relativistic).  This is in good
agreement with the estimated size of the central radio core.  On the
other hand, if the jet velocity is relativistic, the models with
smaller black hole mass would be favored, consistent with the
spectral fitting results.

The above model of the cyclic activity of the central core is based on
the 1-D, vertically averaged calculations and does not describe any
effects in the polar direction. Therefore the jet axis assumed to be
perpendicular to the disk plane is constant in time, and the
subsequent radio structures should be co-aligned. Taking into account,
that the lobe separation is much smaller than the distance to the
source, the viewing angle of $\theta\sim30^{\circ}$ between the line of
sight and the younger radio lobe would be practically the same as the
angle to the older lobe.  In the case of 1045+352 the parts of the jet
we observed in the 5\,GHz structure are not co-aligned. Structures
$A_{1}$ - $A_{2}$ are shifted with respect to the core, while the structure
$A_{3}$ is misaligned by a large angle.

\subsection{Precession of the jet}

The observed misalignment between the young and old
radio structures
could be the result of the changed direction of the jet axis between
the activity episodes. The precession of the jet could be caused by
various mechanisms, such as tidal forces, disk warping,
irradiation, and the Bardeen-Petterson effect \citep{caproni}.  In this
section, we consider the precession of the innermost accretion disk
due to internal instabilities within the accretion flow.  As studied
by \citet{jan08,jan09}, acoustic instabilities of the
Papaloizou-Pringle type can arise in the highly supersonic inner parts
of the accretion flow. The azimuthal modes of such instabilities will
lead to the initially differential tilt of the rotating torus and its
subsequent precession.  The 3-D hydrodynamic simulations were
performed to determine the rate of such precession, which is dependent
on the adopted equation of state (i.e.  the gas adiabatic index), and
scales with the black hole mass.

\begin{figure}
\includegraphics[width=9cm, height=9cm]{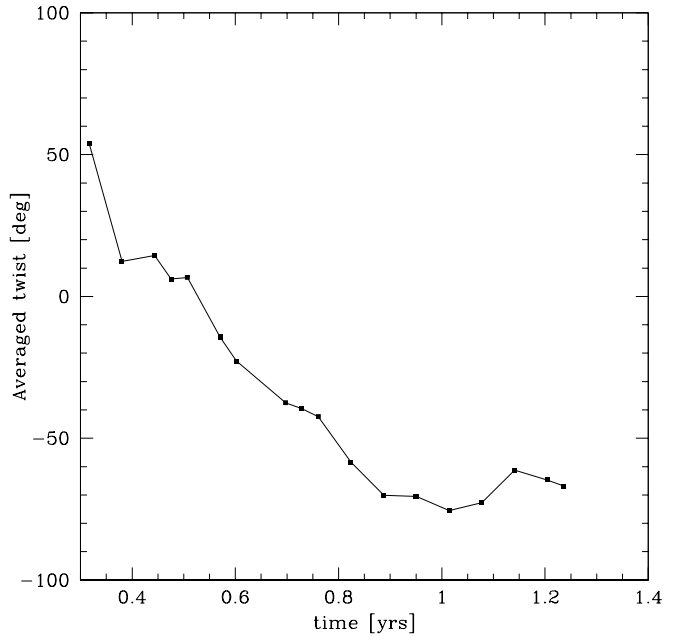}
\caption{The twist angle averaged over radius, from 1.5 to 50 $R_{\rm
Schw}$,
as a function of time. The black hole mass is $2\times 10^{7} M_{\odot}$.}
\label{fig:prec}
\end{figure}

In Figure \ref{fig:prec} we show the precession angle as a function of
time, for $M_{BH}=2\times 10^{7} M_{\odot}$. The simulations were
performed for a few equations of state, and we show here only the
result for the adiabatic index $\gamma=5/3$, which corresponds to the
gas pressure dominated case.
The twist angle, shown in the Figure as a function of time, is
defined as follows:
\begin{equation}
\gamma(r,t) = \arccos\Big({L_{z} \over \sqrt{L_{x}^{2}+L_{y}^{2}}}\Big).
\label{eq:twist}
\end{equation}
where $L_{\rm x}$, $L_{\rm y}$ and $L_{\rm z}$ are the Cartesian
components of the angular momentum vector. The twist is therefore
defined as a cumulative angle by which the angular momentum
vector revolves in the $x-y$ plane by the time $t$.

Initially, when the torus is axisymmetric and no tilt occurs,
the only component of the angular momentum vector is $L_{\rm z}$ and its
sign depends on the direction of the flow rotation. When the torus tilts,
the non-zero $L_{\rm x}$ and $L_{\rm y}$ components
appear and the rotation axis tilts toward the $x-y$ plane.
Initially, the torus precession is differential and the twist angle rises
fast mostly in the innermost parts of the flow. In the Figure
\ref{fig:prec}, we show the twist averaged over radius between the inner 
edge and 50 Schwarzschild radii.
The twist angle decreasing from $\sim 60^{\circ}$
to  $\sim -80^{\circ}$  in the innermost torus
reflects the clockwise precession. The moment, when the precession starts
during the hydrodynamic simulation,
as well as the period of precession, depend on the model (i.e. on the
equation of state). For the model shown
here, we estimated that the precession period is $2.4\times 10^{4}$ in the
units of dynamical time at the inner radius.
For the black hole mass of $2\times 10^{7} M_{\odot}$ it is about 2.5
years

The simulations we described above assume a radiatively inefficient
flow, and therefore they are adequate for the quiescent state of the
central core of the quasar. In other words, the accretion rate in the
precessing disk is negligibly small, and as such is not taken into
account in the 3-D modeling.  Accordingly, we postulate here that the
precession of the disk occurs between the activity episodes, and each
subsequent episode restarts the jet production along a misaligned
axis. The direction of the axis is basically random, because the
period of precession is much smaller (about 20 times) than the
duration of the quiescent phase. In consequence, the younger radio
structure is easily formed at a different angle to the line of sight
than the older one and the probability of that is very high taking
into account the very small precession period.

We note that this approach is a simplification and in order to
really model the precession effect we would need to compute the
3-dimensional model with somewhat
higher accretion rate and radiative efficiency. This is beyond the scope of
our present computations.
However, based on the results of \citet{das04}, we suppose that at least for
not extremely large
accretion rates and viscosities, the effect of precession would be enhanced,
because of the outward
shifting of the sonic point.

\subsection{Interaction with the interstellar medium}

The repetitive outbursts of the central core due to the accretion disk
instability may potentially have a strong impact on the ISM.  Due to
the cyclic irradiation from the central UV source, the shocks in the
ISM may arise on short timescales and keep the gas warm.  The hot
spots can travel to a certain distance within the host galaxy, before
the outburst is finished. This depends on the outburst duration and in
case of a short timescale the radio source does not expand beyond the
host galaxy.  According to \citet{czerny} in order for the source to
escape the host galaxy the outburst would need to be longer than
$10^{4}$ yrs.  For 1045+352, however, this would require a large black
hole mass in this quasar, $M=4\times 10^{8} M_{\odot}$, an accretion
rate equal to the Eddington limit, and a very small viscosity (models A and B
in Fig.~\ref{tab:evol_models}).  Such
values were excluded by our previous spectral modeling
\citep{kunert09}.  In particular, for 1045+352, the estimated core
size is about 70\,pc.

After the jet activity in 1045+352 ceases the radio structure will be
driven further due to the pressure expansion, up to a distance of about
1\,kpc  (for the ISM density of $n=\rho/m_{\rm p}=0.1$ cm$^{-3}$, its
temperature of $10^{7}$ K and the jet power of $L_{\rm jet}= 10^{44}$
erg s$^{-1}$ \citep{kun07}), assuming that the accumulated jet energy is
comparable to the energy content in the heated medium, $E=R^{3}\rho
c_{\rm s}^{2} \sim L_{\rm jet} t$ \citep{stawarz08}.  This distance is
comparable to that of the outermost radio structure, $A_{3}$ in our map
(1.7 kpc). If the jet power was somewhat smaller, the distance of the
radio structures driven by the pressure expansion would also be
smaller, and could be comparable to the inner component $A_{2}$ as well.

When the expansion stops, the radio structure will start recollapsing.
The recollapse time is typically much longer than the outburst cycle,
and the ``tunnel'' made by the expanding jet will not close before the
next outburst.  (For our favorable parameters, the next outburst will
start after about $4\times 10^{3}$ years.)  Therefore the jets will
always propagate into a rarefied medium.  This would imply that the
jet is propagating with a large velocity, and thus the velocity of
$>0.1\,c$ would be favored \citep{stawarz04}.  In this case, any
distortions in the jet morphology would propagate rather quickly and the
apparent twist seen in the maps (i.e. structure $A_{3}$) could be enhanced.
It has to be noted however, that there is a large uncertainty in the
measurements of a black hole mass. According to the SED modeling the
value of the black hole mass in 1045+352 is on the order of $10^{7}
M_{\odot}$, according to the Mg\,II $\lambda$2800\,FWHM it is on the
order of $10^{8} M_{\odot}$.  We therefore cannot exclude the possibility that the
radio structure of 1045+352 will be able to escape from the host
galaxy during this phase of the radio jet activity.

Our analysis of radio components observed in 1045+352 may suggest 
the following interpretation:
the radio jet emerging from the core in a E/SE
direction is not able to get through the dense environment and bends
to NE direction (component $A_{1}$).
The radio emission indicated with $A_{2}$ (Fig.~\ref{1045+352_maps}c)
and visible in a NE direction is probably another trace of
jet-cloud interaction.
Its continuation ($A_{3}$) is visible on the MERLIN 5\,GHz image
(Fig.~\ref{1045+352_maps}b).
The weakness of the counter-jet emission is probably caused by
the large beaming. 
However, the structure may be as well interpreted as involving at least three phases
of quasar activity: components $A_{2}$-$A_{3}$ as the oldest one, structure
$A_{1}$-B as the younger one, and the jet A as the current activity direction.

\subsection{Radio structures of BAL quasars}

Observations based on the VLA FIRST survey have revealed the existence
of radio-loud BAL quasars \citep{becker00, menou01}, which together
with the 5 radio-loud BAL quasars from NVSS discovered by
\citet{broth98} make a large population of objects, about 14\%-18\% of the
total sample of \citet{becker00}.
There are only ten known radio-loud BAL quasars with the extended double-lobed
radio emission \citep{wills99, gregg00, gregg06, broth02, zhou06}.  Among
these are: FR\,II quasar with a steep spectrum core suggesting
restarted activity inside \citep{gregg00}, a hybrid object whose
morphology indicates some type of jet-medium interactions
\citep{wills99}, and a very core-dominated radio triple, probably
beamed source \citep{broth02}. However, most of the BAL quasars remain
unresolved even with the VLBI observations \citep{monte09,doi}. This
implies they are either beamed with the jet pointing close to the line of
sight, or they are very compact objects starting their activity. The resolved ones 
show complex double or triple, in many cases asymmetric, morphology.  
In general, the radio-loud
BALs tend to be compact in the radio, similar to GPS and CSS sources,
which are thought to be the young counterparts of powerful large-scale
radio sources \citep{becker00}. They have wide range of spectral indices
which together with the polarization properties suggest that the radio-loud BAL
quasars are not oriented along a particular line of sight \citep{becker00,
monte09}, contrary to the orientation model. The evolution
interpretation of the nature of BAL phenomena is also still discussed.  

1045+352 is a radio-loud BAL quasar and CSS object with a very complex
structure which suggests that the radio jet is oriented close to the line of sight.
Morphologies like that observed in 1045+352
are seen among other CSSs
\citep[e.g.][]{fanti02, tao}, although they are not very common.
It is very likely that
many CSS objects
interact with an asymmetric medium in the central regions of their
host galaxies, and this can cause the observed asymmetries and
distortions \citep{saikia01, jeyakumar05}. 
In the case of 1045+352 an additional fact may matter, namely, 
thar our {\it Chandra} X-ray observations suggest a presence of the
absorbing material \citep{kunert09} close/in the BLR region. It is then
possible that the
outflowing material detected in absorption may also have interacted
with the radio jet disturbing its structure (see also the interpretation
of the radio morphology of CSS quasar 3C48 by \citet{gupta}).  Such a
strong interaction between a relativistic jet and a BAL wind has been
reported in Mrk\,231 \citep{rey09}, although the interaction takes
place at the small deprojected distance of $\sim4$\,pc from the radio
core. 

Recently \citet{shankar} studied the radio properties of BAL quasars
based on the SDSS/DR3 and FIRST surveys and found that the number of
classical BAL\,QSOs decreases with increasing radio power. They
discussed several plausible physical models which may explain the
observations, namely the evolution and orientation model, however none
of them can fully explain all the features and correlations observed
among radio-loud BAL quasars.  As has already been suggested
\citep{elvis00} not only equatorial disk wind but also cone outflows
can be responsible for BAL features. 
Thus an internal orientation effect such as the opening angle of the accretion
disk wind instead of the external orientation effect such as an angle
between the jet axis and our line of sight, could be what determines a
possibility of observing any BAL features. 

The discussed orientation of 1045+352 is not in agreement with the orientation
model, however, the BAL features are clearly visible in this source. After  
\citet{kunert09} we suggest that there is no a direct dependence between the radio jet
orientation and the probability of BAL visibility.

\section{Summary}
\label{summary}

1045+352 is a CSS object and a HiBAL quasar at a medium redshift.  The
spectrum classification and linear size of the source indicate that it is a
young object in the early phase of its evolution.  The new more
sensitive high-resolution EVN+MERLIN 5\,GHz observations revealed details of
the source radio emission and confirmed its complicated
radio structure. The radio morphology of 1045+352 is dominated by the
strong radio jet resolved into many subcomponents and changing its
orientation during propagation in the central regions of the host
galaxy.
The presence of the dense environment and material outflows have been
already confirmed in this source based on the submillimeter and X-ray
observations and as we discussed, they can have a significant impact
on the radio jet morphology of 1045+352. We conclude that the most
probable interpretation of the observed radio structure of 1045+352 is
the ongoing process of the jet precession due to internal
instabilities within the flow and/or jet interaction with a dense,
inhomogeneous interstellar medium.

It is difficult to establish the orientation between the jet axis and
the observer in 1045+352 because of the complex structure.
Nevertheless taking into account the innermost radio structure
we suggest that the radio jet is oriented close to the line of sight. This means
that the opening angle of the accretion disk wind can be
large in this source.

\acknowledgements
We thank Peter Thomasson for his help with
MERLIN observations and data reduction.
MERLIN is a UK National Facility operated by
the University of Manchester on behalf of STFC.

We thank Zsolt Paragi and Bob Campbell for their help with
EVN data reduction. The European VLBI Network is a joint facility of
European, Chinese, South African and other radio astronomy institutes funded
by their national research councils.

This work was supported by the Polish Ministry of Science and
Higher Education under grant N N203 303635.

\end{document}